\documentclass[aps,preprint,prd,showpacs,nofootinbib]{revtex4}

\usepackage{latexsym}
\usepackage{amsmath}
\usepackage{graphicx}
\usepackage{subfigure}
\usepackage{dcolumn}
\usepackage{bm}
\usepackage{amssymb}
\usepackage{latexsym}
\usepackage{ulem}
\usepackage{color}
\usepackage[colorlinks,linkcolor=magenta,anchorcolor=cyan,citecolor=blue]{hyperref}

\def\be{\begin{equation}}
\def\ee{\end{equation}}
\def\ba{\begin{eqnarray}}
\def\ea{\end{eqnarray}}

\def\nn{\nonumber}
\def\lf{\left}
\def\rt{\right}

\begin{document}

\title{A covariant Lagrangian for stable nonsingular bounce }

\author{Yong Cai$^{1}$\footnote{caiyong13@mails.ucas.ac.cn}}
\author{Yun-Song Piao$^{1,2}$\footnote{yspiao@ucas.ac.cn}}

\affiliation{$^1$ School of Physics, University of Chinese Academy of
Sciences, Beijing 100049, China}

\affiliation{$^2$ Institute of Theoretical Physics, Chinese
Academy of Sciences, P.O. Box 2735, Beijing 100190, China}

\begin{abstract}


The nonsingular bounce models usually suffer from the ghost or
gradient instabilities, as has been proved recently. In this
paper, we propose a covariant effective theory for stable
nonsingular bounce, which has the quadratic order of the second
order derivative of the field $\phi$ but the background set only
by $P(\phi,X)$. With it, we explicitly construct a fully stable
nonsingular bounce model for the ekpyrotic scenario.

\end{abstract}

\maketitle

\section{Introduction}

General relativity (GR) suffers the singularity problem
\cite{Hawking:1969sw}, which indicates the incompleteness of our
understanding about the gravity theory as well as the origin of
the Universe \cite{Borde:1993xh}\cite{Borde:2001nh}. Instead of looking for a UV(ultraviolet)-complete
theory to describe what happens at the ``singularity",
investigating the possibility of a nonsingular origin of the
Universe with the effective theory, which captures low energy
behaviors of the complete theory, is a significant direction.

It seems that since \cite{Cai:2007qw}, the perturbations of the Friedmann-Roberson-Walker background usually suffer from the ghost or gradient
instabilities in nonsingular
cosmological models, see \cite{Rubakov:2014jja} for a review.
Recently,
this observation has been proved, up to the cubic Galileon theory
\cite{Libanov:2016kfc} and the Horndeski theory
\cite{Kobayashi:2016xpl}. Based on the effective field theory
(EFT) of nonsingular cosmologies
\cite{Cai:2016thi}\cite{Creminelli:2016zwa}\cite{Cai:2017tku},
this No-go result has been more clearly illustrated. It is found
that the stable nonsingular cosmological models can be implemented
only in the theories beyond cubic Galileon, (see also
\cite{Kolevatov:2016ppi}\cite{Akama:2017jsa}).

Recent progresses have inspired a wave of looking for stable
nonsingular bounce
\cite{Ijjas:2016vtq}\cite{deRham:2017aoj}\cite{Yoshida:2017swb}
(see also \cite{Misonoh:2016btv}\cite{Giovannini:2016jkf}), along
the road beyond the cubic Galileon (even the Horndeski theory
\cite{Horndeski:1974wa}\cite{Deffayet:2009mn}\cite{Kobayashi:2011nu}).
Moreover, the developments of scalar-tensor theory (the GLPV
\cite{Gleyzes:2014dya} and DHOST theory
\cite{Langlois:2015cwa}\cite{Langlois:2015skt}\cite{Langlois:2017mxy},
the mimetic gravity
\cite{Chamseddine:2014vna}\cite{Chamseddine:2016uef}) might also
be able to provide us with some chances to implement stable
nonsingular cosmologies. However, due to the complexity of
relevant theories, which component is required for a stable bounce
is not clear. Thus so far building a realistic and stable model is
still difficult.

In Refs.\cite{Cai:2016thi}\cite{Creminelli:2016zwa}, with the EFT
of nonsingular cosmologies, it has been found that the operator
$R^{(3)}\delta g^{00}$ is significant for the stability of
nonsingular bounce. Actually, in unitary gauge, without getting
involved in the specific theories, \be L_{add-oper}\sim
{M_2^4(t)\over2}(\delta g^{00})^2 +{\tilde{m}_4^2(t)\over
2}R^{(3)}\delta g^{00}\label{EFT}\ee might be the least set of
operators added to GR to cure the instabilities, since $(\delta
g^{00})^2\sim {\dot \zeta}^2$ while $R^{(3)}\delta g^{00}\sim
({\partial \zeta})^2$ at quadratic order.

In this paper, based on the covariant description of the
$R^{(3)}\delta g^{00}$ operator, we propose a covariant theory for
stable nonsingular bounce, which has the quadratic order of
the second order derivative of the field $\phi$ but the background
set only by $P(\phi,X)$.
We illuminate its
application by constructing a fully stable nonsingular bounce
model for the ekpyrotic scenario
\cite{Khoury:2001wf}\cite{Lehners:2008vx}.

\textbf{Note added:} Several days after our paper appeared in arXiv, the preprint \cite{Kolevatov:2017voe} appeared, in which somewhat similar analysis is done in beyond Horndeski model with sort of similar result.


\section{Covariant description of $R^{(3)}\delta g^{00}$ }

In unitary gauge, $\phi=\phi(t)$. We have \be \delta
g^{00}={X\over \dot{\phi}^2(t)}+1={X\over f_2(t(\phi))}+1,
\label{g00}\ee  where $X=\phi_{\mu} \phi^{\mu}$,
$\phi_\mu=\nabla_\mu\phi$ and $\phi^\mu=\nabla^\mu\phi$.

$R^{(3)}$ is the Ricci scalar on the 3-dimensional spacelike
hypersurface. Using the Gauss-Codazzi relation, it is
straightforward (though tedious) to find \ba \label{covaR3}
R^{(3)}&=& R-{\phi_{\mu\nu}\phi^{\mu\nu}-(\Box \phi)^2\over X}
+{2\phi^\mu\phi_{\mu\nu}\phi^{\nu\sigma}\phi_\sigma \over
X^2}-{2\phi^\mu \phi_{\mu\nu}\phi^\nu \Box \phi\over X^2}
\nn\\&\,& +{2(\phi^\nu_{~\nu\mu}\phi^\mu -\phi_{\nu~\mu}^{~\mu~}
\phi^\nu)\over X}\,, \ea with
$\phi_{\mu\nu}=\nabla_\nu\nabla_\mu\phi$ and
$\phi^\nu_{~\nu\mu}=\nabla_\mu\nabla_\nu \nabla^\nu\phi$. It is
simple to check that the right hand side of Eq. (\ref{covaR3}) is
0 at the background level.


We define $S_{\delta g^{00} R^{(3)}}=\int d^4x\sqrt{-g} L_{\delta
g^{00} R^{(3)}}$, and have
\ba \label{covaaction} L_{\delta g^{00} R^{(3)}}& =& {f_1(\phi)\over 2}\delta g^{00} R^{(3)}\nn\\
&=& {f\over 2}R - {X\over2}\int f_{\phi\phi}d \ln X
-\lf(f_\phi+\int {f_\phi \over 2}d\ln X\rt)\Box\phi \nn\\
& + & {f\over 2X}\lf[\phi_{\mu\nu}\phi^{\mu\nu}-(\Box
\phi)^2\rt]-{f-2Xf_X\over
X^2}\lf[\phi^\mu\phi_{\mu\rho}\phi^{\rho\nu}\phi_\nu-(\Box
\phi)\phi^\mu\phi_{\mu\nu}\phi^\nu\rt]
\ea   after integration by parts, where
$f(\phi,X)=f_1\lf(1+{X\over f_2}\rt)$ has the dimension of mass squared, $f_2(\phi)$ is defined in
(\ref{g00}), and the total derivative terms have been
discarded. One useful formula for obtaining Eq.
(\ref{covaaction}) is \be 2{\cal
B}(\phi,X)\phi^\mu\phi_{\mu\nu}\phi^\nu = \nabla_\mu\lf(\phi^\mu
\int {\cal B}dX\rt)- X\int {\partial{\cal B}\over \partial \phi}dX
-\Box\phi\int {\cal B}dX\,. \ee

\section{Stable nonsingular bounce}

\subsection{The covariant theory}

Here, the EFT proposed is \be S=\int
d^4x\sqrt{-g}\lf({M_p^2\over2}R +P(\phi,X)\rt)+S_{\delta g^{00}
R^{(3)}}\,,\label{action01}\ee which is a covariant theory
equivalent to GR plus the set of operators in (\ref{EFT}),
since $M_2^4(t)=\dot{\phi}^4P_{XX}$ and ${\tilde
m}_4^2(t)=f_1(\phi)$.

The covariant action (\ref{action01}) actually belongs to a
subclass of the DHOST theory
\cite{Langlois:2015cwa}\cite{Langlois:2015skt} (see Appendix \ref{App} for details), which could avoid
the Ostrogradski instability, up to quadratic order of the second
order derivative of $\phi$. Ijjas and Steinhardt used the quartic
Horndeski action in \cite{Ijjas:2016vtq}. In (\ref{covaaction}),
though the nonminimal coupling $f(\phi,X)R$ is similar to that in
\cite{Ijjas:2016vtq}, terms $\sim\Box \phi$,
$\phi_{\mu\nu}\phi^{\mu\nu}$, $(\Box \phi)^2$, $(\Box
\phi)\phi^\mu\phi_{\mu\nu}\phi^\nu$ and
$\phi^\mu\phi_{\mu\rho}\phi^{\rho\nu}\phi_\nu$ also appear
simultaneously with the coefficients set by $\delta g^{00}
R^{(3)}$, so that the effect of $S_{\delta g^{00} R^{(3)}}$ on
background is canceled accurately.
Here, the background is set
only by $P(\phi,X)$. In \cite{deRham:2017aoj}, $(\Box\phi)^2$ is
used, which shows itself the Ostrogradski ghost, see also earlier
\cite{Li:2005fm}, how to remove it requires argumentation.


The quadratic action of scalar perturbation for (\ref{action01})
is \be S_{\zeta}^{(2)}=\int a^3 Q_s\lf(\dot{\zeta}^2-c_s^2
{(\partial\zeta)^2\over a^2} \rt)d^4x \,,\label{scalar-action} \ee
in which \be Q_s={2{\dot \phi}^4P_{XX}-M_p^2{\dot H}\over H^2},\quad
c_s^2Q_s=M_p^2\lf({{\dot c}_3\over a} -1\rt)\label{cs2} \ee  and
$c_3=a(1+{2f_1\over M_p^2})/H$.  We can see that the sound speed of scalar perturbation can be directly modified by $f_1(\phi)$, namely, the function before $\delta g^{00}R^{(3)}$ operator. Therefore, the gradient instability of scalar perturbation could be cured by proper choice of $f_1(\phi)$, while that of tensor perturbation
is unaffected by $S_{\delta g^{00} R^{(3)}}$, hence is same with
that of GR.

A fully stable nonsingular bounce ($Q_s>0$ and $c_s^2=1$) can be
designed with (\ref{action01}). In the bounce phase, ${\dot H}>0$.
However, $Q_s>0$ can be obtained, since $P(\phi, X)$ contributes
$\dot{\phi}^4P_{XX}$ in $Q_s$. While around the bounce point
$H\simeq 0$, \be c_s^2\sim -{\dot H}\lf(1+{2f_1\over M_p^2}\rt).
\ee Thus we will have $c_s^2>0$ for $2f_1<-{M_p^2}$, as has been
clarified in Refs.\cite{Cai:2016thi}\cite{Cai:2017tku}. It should
be mentioned that if $f_1=0$, we have $c_s^2\sim -{\dot H}<0$
around the bounce point, thus $S_{\delta g^{00} R^{(3)}}$ is
needed to contribute $f_1$. Here, we always could set $c_s^2\sim
{\cal O}(1)$ with a suitable $f_1(\phi)$  (see also
\cite{Cai:2017tku}) which satisfies  \be {2f_1(\phi)}={ H\over
a}\int a\lf(Q_s c_s^2 +M_p^2\rt)dt-M_p^2. \label{m4tilde}\ee




\subsection{A stable nonsingular bounce model}\label{pbounceinf}

With (\ref{action01}), building a nonsingular bounce model is
simple. The ghost-free nonsingular bounce is set by $P(\phi, X)$,
while $c_s^2\simeq 1$ is set by using suitable $f_1$ and $f_2$ in
(\ref{g00}).

As a specific model, we set $P(\phi,X)$ in (\ref{action01})
as \be P(\phi,X)= \lf[{k_0\over (1+\kappa_1\phi^2)^2}-1\rt] {X/2}+
{q_0\over (1+\kappa_2\phi^2)^2 }X^2-V(\phi)\,,\label{P} \ee where
the potential is ekpyrotic-like
\begin{eqnarray} \label{vphi} &&V(\phi)=-{ V_0\over 2} e^{\phi/{\cal M}_1
}\left[1-\tanh( {\phi\over {\cal M}_2})\right] \,,
\end{eqnarray} with constant ${\cal M}_1, {\cal M}_2, V_0$,
and $k_0, \kappa_1$ responsible for the switching of the sign
before $X/2$ around $\phi\simeq 0$, and $q_0, \kappa_2$ for the
appearance of $X^2$ around $\phi\simeq 0$, see
\cite{Koehn:2015vvy} for a similar $P(\phi,X)$, which might allow
for a supersymmetric counterpart \cite{Koehn:2013upa}.

The background equations are \ba 3 M_p^2H^2  & =&
-2\dot{\phi}^2P_X-P\,,\\
M_p^2\dot{H}  & =& \dot{\phi }^2 P_X \,.\label{dotH}\label{eomphi}
\ea Initially $\phi\ll -{\cal M}_2, -1/\sqrt{\kappa_1},
-1/\sqrt{\kappa_2}$, we have $P(\phi,X)=-X/2{ +V_0 e^{\phi/{\cal
M}_1 } }$, the Universe is in the ekpyrotic phase with the equation
of state parameter\be \omega_{ekpy}={ M_p^2\over 3{\cal
M}_1^2}-1>1. \ee

Around $\phi\simeq 0$, we have \be   {\dot H}\simeq \lf({k_0-1\over
2}-2q_0{\dot \phi}^2\rt){\dot \phi}^2>0. \ee Thus the bounce could
occur. However, after the bounce the field $\phi$ will be
canonical again but with $V(\phi)=0$. It is possible that the
phase after the bounce might be the inflation \cite{Piao:2003zm}\cite{Liu:2013kea}, we will consider it
elsewhere.

Here, in the quadratic action (\ref{scalar-action}) of scalar
perturbation, \be Q_s= -{M_p^2{\dot H}\over H^2}+{4q_0\over
(1+\kappa_2\phi^2)^2 H^2}\dot{\phi}^4>0 \ee can be obtained,
while $c_s^2=1$ can be obtained by setting suitable $f_1(\phi)$
in (\ref{covaaction}), which is given by (\ref{m4tilde}), and
$f_2(\phi)={\dot \phi}(t(\phi))$.

\begin{figure}[htbp]
\subfigure[~~$\phi$]{\includegraphics[width=.47\textwidth]{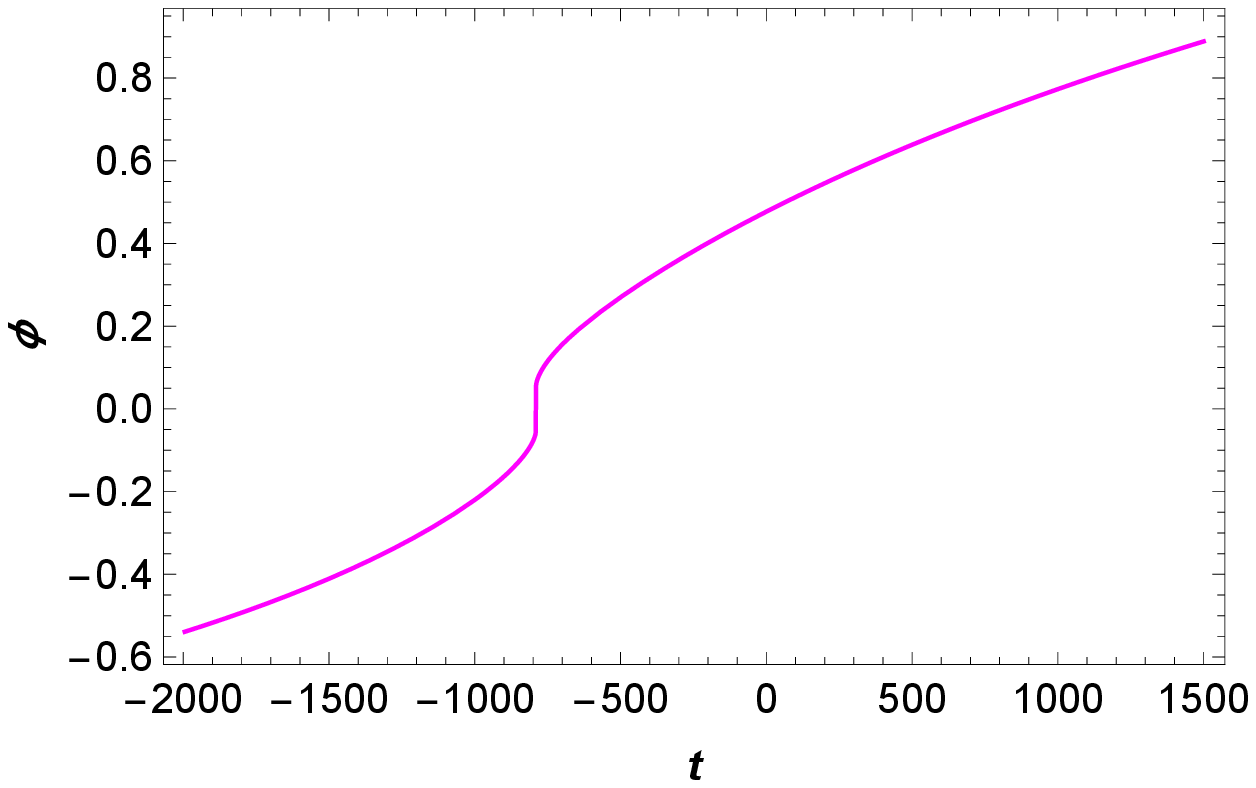}
}
\subfigure[~~$a$]{\includegraphics[width=.45\textwidth]{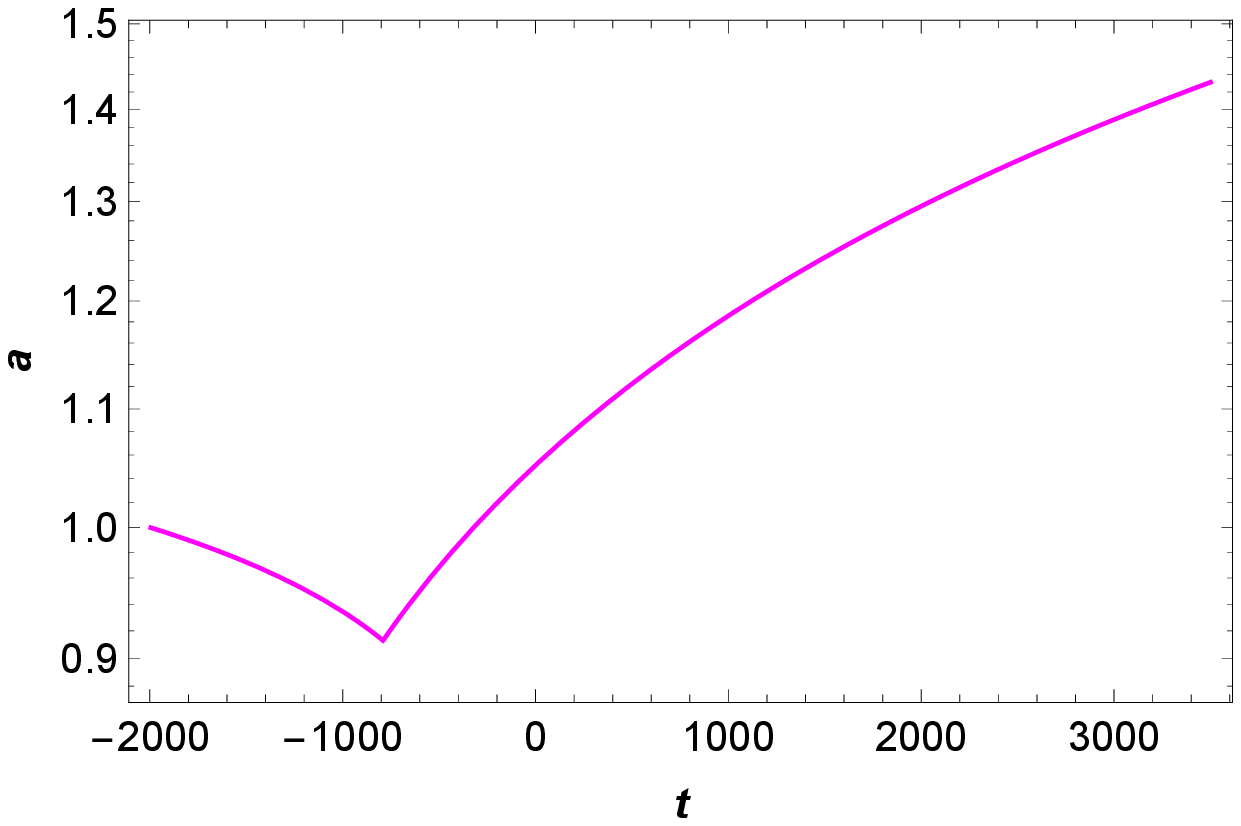}
}
\subfigure[~~$H$]{\includegraphics[width=.47\textwidth]{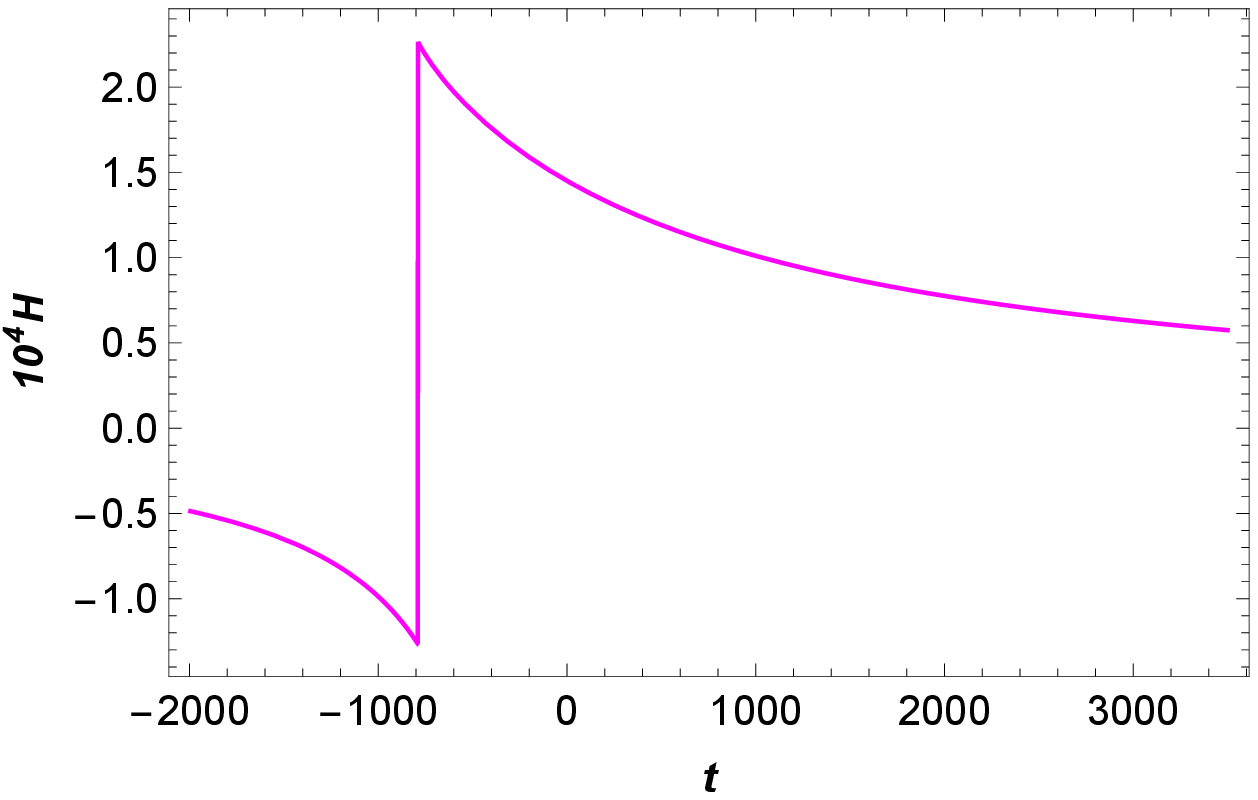}
}
\subfigure[~~$\epsilon=-\dot{H}/H^2$]{\includegraphics[width=.46\textwidth]{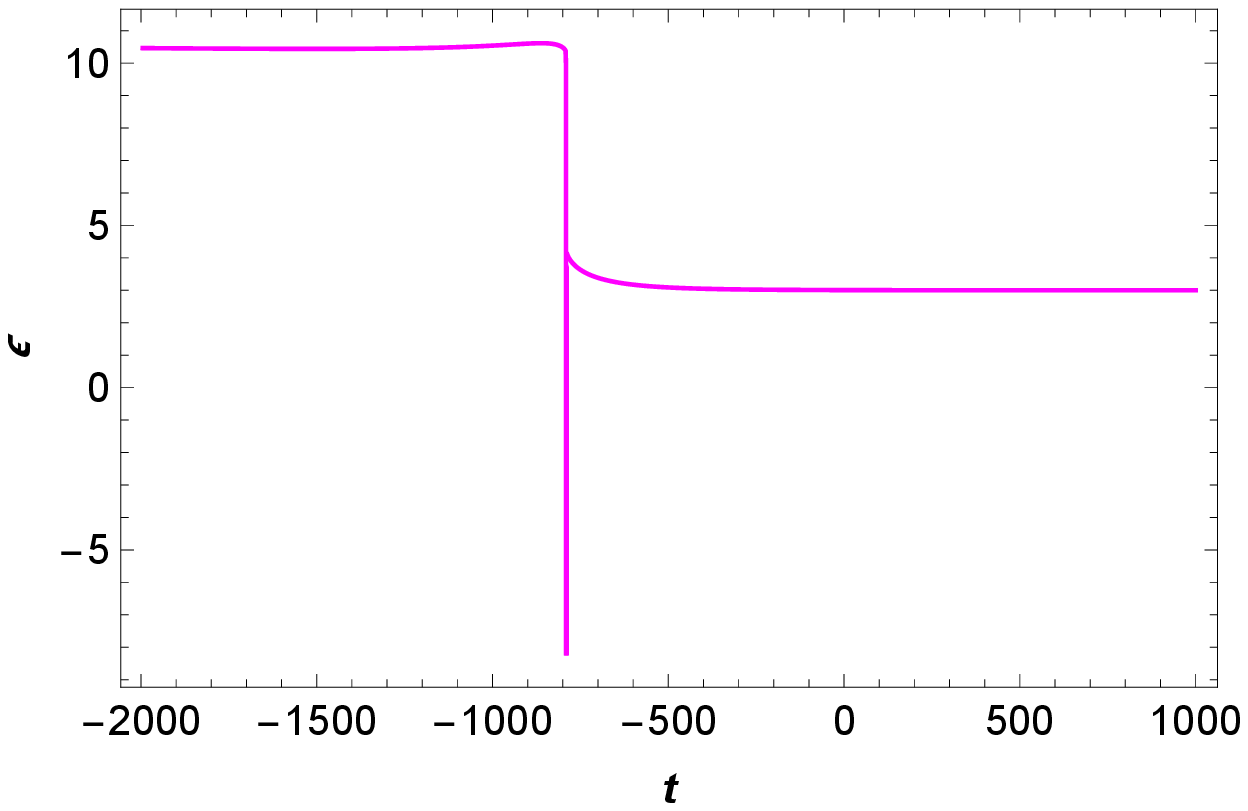}
} \caption{The background evolution of ekpyrotic Universe.}
\label{fig01}
\end{figure}

\begin{figure}[htbp]
\subfigure[~~$f_1(\phi)$ for $c_s^2\equiv
1$]{\includegraphics[width=.445\textwidth]{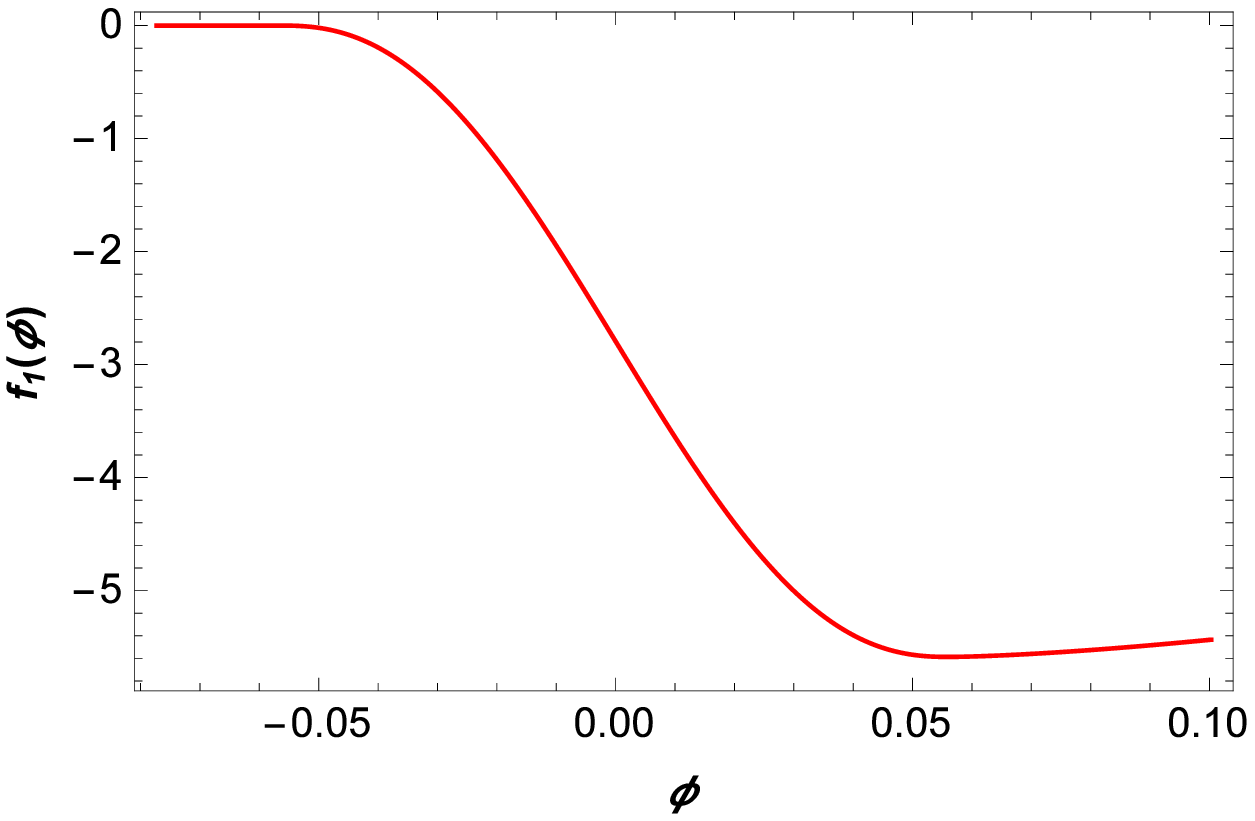} }
\subfigure[~~$f_2(\phi)$ for $c_s^2\equiv
1$]{\includegraphics[width=.47\textwidth]{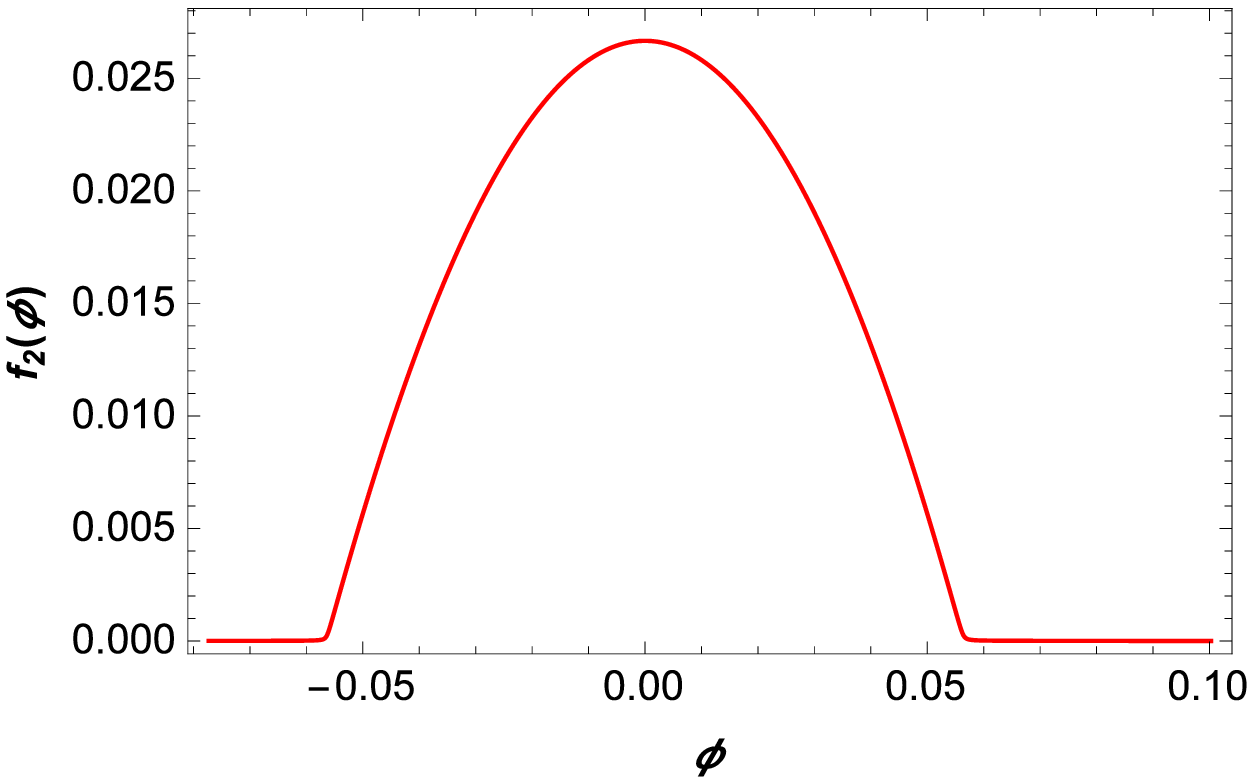} }
\caption{The expressions of $f_1(\phi)$ and $f_2(\phi)$ with
respect to $\phi$.} \label{fig02}
\end{figure}

The background evolution is numerically plotted in
Fig. \ref{fig01}. We show the behaviors of $f_1(\phi)$ and
$f_2(\phi)$ with respect to $\phi$ in Fig. \ref{fig02} while we
require $c_s^2=1$ throughout. In both Figs. \ref{fig01} and
\ref{fig02}, we set $k_0=1.2$, $\kappa_1=30$, $q_0=1.25$,
$\kappa_2=20$, $V_0=2\times 10^{-7}$, ${\cal M}_1=0.22$ and ${\cal
M}_2=0.1$. We set the initial condition of $\phi$ as
$\phi_{ini}=-0.54$ and $\dot{\phi}_{ini}=2.24\times 10^{-4}$, while
the initial value of $t$ is $t_{ini}=-2000$. We see that with
$f_1$ and $f_2$ plotted in Fig. \ref{fig02}, the Lagrangian
(\ref{action01}) with $P(\phi,X)$ in (\ref{P}) will bring a fully
stable nonsingular bounce ($Q_s>0$ and $c_s^2=1$).

\section{Discussion}

The exploration of stable nonsingular bounce has been still a
significant issue. Recently, it has been found in
Refs.\cite{Cai:2016thi}\cite{Creminelli:2016zwa} that the operator
$R^{(3)}\delta g^{00}$ in EFT of nonsingular cosmologies is
significant for the stability of bounce. Here, based on the
covariant description of the $R^{(3)}\delta g^{00}$ operator, we
propose a covariant theory (\ref{action01}) for stable nonsingular
bounce.


Our (\ref{action01}) is actually a subclass of the DHOST theory
\cite{Langlois:2015cwa}\cite{Langlois:2015skt}, but the
cosmological background is set only by $P(\phi,X)$. The
$P(\phi,X)$ nonsingular bounce model could be ghost-free
\cite{Buchbinder:2007ad}\cite{Koehn:2015vvy}, but suffers the
problem of $c_s^2<0$, which can not be dispelled by using the
Galileon interaction $\sim \Box \phi$
\cite{Libanov:2016kfc}\cite{Kobayashi:2016xpl}\cite{Cai:2016thi}\cite{Creminelli:2016zwa}.
Actually, in \cite{Ijjas:2016tpn}\cite{Cai:2017tku}, it is
observed that the Galileon interaction only moves the period of
$c_s^2<0$ to the outside of the bounce phase, but can not remove
it, see also earlier \cite{Easson:2011zy}.  Thus it could be
imagined that the quadratic order of the second order derivative
of $\phi$, i.e., $\phi_{\mu\nu}\phi^{\mu\nu}$, $(\Box \phi)^2$,
$\phi^\mu\phi_{\mu\rho}\phi^{\rho\nu}\phi_\nu$ and $(\Box
\phi)\phi^\mu\phi_{\mu\nu}\phi^\nu$, might play crucial roles in
stable nonsingular bounce model. However, due to the complexity of
relevant theories, what kind of combination of these components is
required for a stable cosmological bounce is unclear. Here, the
corresponding combination (\ref{covaaction}) is just what told by
the covariant description of the $R^{(3)}\delta g^{00}$ operator.

With (\ref{action01}), the design of stable nonsingular bounce
model is simple, as illuminated for the ekpyrotic scenario. Our
work actually offers a concise way to the fully stable nonsingular
cosmologies.
See also \cite{Biswas:2011ar}\cite{Odintsov:2014gea}\cite{Banerjee:2016hom}\cite{Hendi:2016tiy} for other interesting studies.

Here, the importance of the EFT of nonsingular cosmologies is
obvious. Actually, the role of $R^{(3)}\delta K$ in EFT
\cite{Cai:2016thi} is similar to that of $R^{(3)}\delta g^{00}$,
where $K_{\mu\nu}$ is the extrinsic curvature on the 3-dimensional
spacelike hypersurfaces. The covariant description of
$R^{(3)}\delta K$ involves the term $\sim (\Box\phi)R$, which
might have the Ostrogradski ghost unless certain constraint is
imposed. This issue will be revisited. In mimetic gravity
\cite{Chamseddine:2014vna}\cite{Chamseddine:2016uef} (see e.g. \cite{Sebastiani:2016ras} for review), since the
mimetic constraint suggests $\delta g^{00}=0$ (which is the
source of instabilities
\cite{Cognola:2016gjy}\cite{Ijjas:2016pad}\cite{Firouzjahi:2017txv}\cite{Hirano:2017zox}),
one might apply the operator $R^{(3)}\delta K$ to make the
(possibly-built) nonsingular bounce stable \footnote{Communication
with Mingzhe Li.}, instead of $R^{(3)}\delta g^{00}$. The mimetic
gravity with the couple $(\Box\phi)R$ has been proposed in
Ref.\cite{Zheng:2017qfs}. We will back to the relevant issues.


\textbf{Acknowledgments}

We thank Mingzhe Li, Taotao Qiu and Youping Wan for helpful discussions. This
work is supported by NSFC, No. 11575188, 11690021, and also
supported by the Strategic Priority Research Program of CAS, No.
XDA04075000, XDB23010100.

\appendix

\section{Correspondence with a subclass of DHOST theory }\label{App}

Up to cubic order of $\phi_{\mu\nu}$, the covariant action  of DHOST can be written as (see e.g., \cite{Langlois:2017mxy})
\ba \label{covaDHODST}
S_{DHOST} &=& \int d^4 x \sqrt{- g }
\left[ p(\phi,X) + q(\phi,X) \Box \phi
+
g_2(\phi,X)  R + C_{(2)}^{\mu\nu\rho\sigma}   \phi_{\mu\nu} \phi_{\rho\sigma}
\right.
\nn\\
&&
\left. \qquad \qquad\qquad + g_3(\phi,X) G_{\mu\nu} \phi^{\mu\nu}  +  
C_{(3)}^{\mu\nu\rho\sigma\alpha\beta}  \phi_{\mu\nu} \phi_{\rho\sigma} \phi_{\alpha \beta} \right]  \,,
\ea
where $R$ and $G_{\mu\nu}$ denote the usual 4-dimensional Ricci scalar and Einstein tensor associated with the metric $g_{\mu\nu}$, respectively;
\be
\label{C2}
C_{(2)}^{\mu\nu\rho\sigma}   \phi_{\mu\nu}  \phi_{\rho\sigma} =\sum_{A=1}^{5}a_A(\phi,X)\,   L^{(2)}_ A\,,
\ee
with 
\be
\label{QuadraticL}
\begin{split}
	& L^{(2)}_1 = \phi_{\mu \nu} \phi^{\mu \nu} \,, \qquad
	L^{(2)}_2 =(\Box \phi)^2 \,, \qquad
	L_3^{(2)} = (\Box \phi) \phi^{\mu} \phi_{\mu \nu} \phi^{\nu} \,,  \\
	& L^{(2)}_4 =\phi^{\mu} \phi_{\mu \rho} \phi^{\rho \nu} \phi_{\nu} \,, \qquad
	L^{(2)}_5= (\phi^{\mu} \phi_{\mu \nu} \phi^{\nu})^2\,,
\end{split}
\ee
and
\be
\label{C3}
C_{(3)}^{\mu\nu\rho\sigma\alpha\beta} \phi_{\mu\nu}  \phi_{\rho\sigma}  \phi_{\alpha \beta} = \sum_{A=1}^{10} b_A(\phi,X)  L^{(3)}_A \,,
\ee
with 
\be
\label{CubicL}
\begin{split}
	& L^{(3)}_1=  (\Box \phi)^3  \,, \quad
	L^{(3)}_2 =  (\Box \phi) \phi_{\mu \nu} \phi^{\mu \nu} \,, \quad
	L^{(3)}_3= \phi_{\mu \nu}\phi^{\nu \rho} \phi^{\mu}_{\rho} \,,   \\
	& L^{(3)}_4= \left(\Box \phi\right)^2 \phi_{\mu} \phi^{\mu \nu} \phi_{\nu} \,, \quad
	L^{(3)}_5 =  \Box \phi\, \phi_{\mu}  \phi^{\mu \nu} \phi_{\nu \rho} \phi^{\rho} \,, \quad
	L^{(3)}_6 = \phi_{\mu \nu} \phi^{\mu \nu} \phi_{\rho} \phi^{\rho \sigma} \phi_{\sigma} \,,   \\
	& L^{(3)}_7 = \phi_{\mu} \phi^{\mu \nu} \phi_{\nu \rho} \phi^{\rho \sigma} \phi_{\sigma} \,, \quad
	L^{(3)}_8 = \phi_{\mu}  \phi^{\mu \nu} \phi_{\nu \rho} \phi^{\rho}\, \phi_{\sigma} \phi^{\sigma \lambda} \phi_{\lambda} \,,   \\
	& L^{(3)}_9 = \Box \phi \left(\phi_{\mu} \phi^{\mu \nu} \phi_{\nu}\right)^2  \,, \quad
	L^{(3)}_{10} = \left(\phi_{\mu} \phi^{\mu \nu} \phi_{\nu}\right)^3 \,;
\end{split}
\ee
extra conditions on the functions $a_A$ and $b_A$ need to be satisfied so that there is no extra propagating degree of freedom, see \cite{Langlois:2017mxy} and references therein for further discussions.

Comparing with (\ref{covaDHODST}), we find our model (\ref{action01})  corresponds to the covariant form of DHOST theory with
\ba &\,&
p(\phi,X)=P(\phi,X)- {X\over2}\int f_{\phi\phi}d \ln X\,,
\qquad q(\phi,X)=-f_\phi-\int {f_\phi \over 2}d\ln X\,,
\nn\\&\,&
g_2(\phi,X)={M_p^2+f\over 2}\,,
\qquad
g_3(\phi,X)=0\,,
\nn\\&\,&
a_1=-a_2={f\over2X}\,,\qquad
a_3=-a_4={f-2Xf_X\over
	X^2} \,,
\qquad
a_5= 0\,,\label{ab000}
\ea
and $b_A=0$.

In the EFT formalism,
the quadratic action for DHOST theory can be written as
\ba
\label{DHOST}
S^{(2)}_{DHOST} &=& \int d^3x \,  dt \,  a^3  \frac{M^2}2\bigg\{
\delta K_{\mu\nu}\delta K^{\mu\nu}
-\lf(1+{2\over3}\alpha_L \rt)\delta K^2
+(1+\alpha_T)\lf(R^{(3)}{\delta \sqrt{h}\over a^3}+\delta_2 R^{(3)} \rt)
\nn\\
&\,&+H^2\alpha_K \delta N^2
+4H\alpha_B \delta K\delta N
+(1+\alpha_H)R^{(3)}\delta N
+4\beta_1\delta K \delta \dot{N}
+\beta_2 \delta \dot{N}^2
+{\beta_3\over a^2}(\partial_i \delta N)^2
\bigg\} \; ,
\nn\\
\ea
where $\delta N=\delta g^{00}/2$, $\delta_2 R^{(3)}$ stands for the second order term in the perturbative expansion of $R^{(3)}$, the dimensionless time-dependent functions $\alpha_L$, $\alpha_T$,  $\alpha_K$, $\alpha_B$, $\alpha_H$, $\beta_1$, $\beta_2$ and $\beta_3$ satisfy certain conditions so that there is no extra propagating degree of freedom, see \cite{Langlois:2017mxy} for details.

Comparing with (\ref{DHOST}), we find
our model (\ref{action01})  corresponds to 
\ba &\,&
M=M_p\,,\qquad \alpha_L=\alpha_T=\alpha_B=0\,,
\qquad \beta_1=\beta_2=\beta_3=0\,,
\nn\\&\,&
\alpha_K={4M_2^4\over M_p^2H^2}={4X^2P_{XX}\over M_p^2 H^2}\,,\qquad
\alpha_H={2\tilde{m}_4^2\over M_p^2}={2f_1(\phi)\over M_p^2}
\,.\label{corresp}
\ea
Thus our model (\ref{action01}) belongs
to a subclass of the DHOST theory.
Note that the results in
Eqs. (\ref{corresp}) should be evaluated at background level in the quadratic action if we derive them from Eqs. (\ref{ab000}) by using formulae given in Eqs. (2.14) of \cite{Langlois:2017mxy}.


\begin{thebibliography}{99}

\bibitem{Hawking:1969sw}
  S.~W.~Hawking and R.~Penrose,
  Proc.\ Roy.\ Soc.\ Lond.\ A {\bf 314}, 529 (1970).


\bibitem{Borde:1993xh}
  A.~Borde and A.~Vilenkin,
  Phys.\ Rev.\ Lett.\  {\bf 72}, 3305 (1994)
  [gr-qc/9312022].

\bibitem{Borde:2001nh}
  A.~Borde, A.~H.~Guth and A.~Vilenkin,
  Phys.\ Rev.\ Lett.\  {\bf 90}, 151301 (2003)
  [gr-qc/0110012].

\bibitem{Cai:2007qw}
  Y.~F.~Cai, T.~Qiu, Y.~S.~Piao, M.~Li and X.~Zhang,
  JHEP {\bf 0710}, 071 (2007)
  [arXiv:0704.1090 [gr-qc]].

\bibitem{Rubakov:2014jja}
  V.~A.~Rubakov,
  Phys.\ Usp.\  {\bf 57}, 128 (2014)
  [Usp.\ Fiz.\ Nauk {\bf 184}, 2, 137 (2014)]
  [arXiv:1401.4024 [hep-th]].


\bibitem{Libanov:2016kfc}
  M.~Libanov, S.~Mironov and V.~Rubakov,
  JCAP {\bf 1608}, 08, 037 (2016)
  [arXiv:1605.05992 [hep-th]].


\bibitem{Kobayashi:2016xpl}
  T.~Kobayashi,
  Phys.\ Rev.\ D {\bf 94}, 4, 043511 (2016)
  [arXiv:1606.05831 [hep-th]].


\bibitem{Cai:2016thi}
  Y.~Cai, Y.~Wan, H.~G.~Li, T.~Qiu and Y.~S.~Piao,
  JHEP {\bf 1701}, 090 (2017)
  [arXiv:1610.03400 [gr-qc]].

\bibitem{Creminelli:2016zwa}
  P.~Creminelli, D.~Pirtskhalava, L.~Santoni and E.~Trincherini,
  JCAP {\bf 1611}, 11, 047 (2016)
  [arXiv:1610.04207 [hep-th]].


\bibitem{Cai:2017tku}
  Y.~Cai, H.~G.~Li, T.~Qiu and Y.~S.~Piao,
  arXiv:1701.04330 [gr-qc].



\bibitem{Kolevatov:2016ppi}
  R.~Kolevatov and S.~Mironov,
  Phys.\ Rev.\ D {\bf 94}, 12, 123516 (2016)
  [arXiv:1607.04099 [hep-th]].



\bibitem{Akama:2017jsa}
  S.~Akama and T.~Kobayashi,
  Phys.\ Rev.\ D {\bf 95}, 6, 064011 (2017)
  [arXiv:1701.02926 [hep-th]].




\bibitem{Ijjas:2016vtq}
  A.~Ijjas and P.~J.~Steinhardt,
  Phys.\ Lett.\ B {\bf 764}, 289 (2017)
  [arXiv:1609.01253 [gr-qc]].



\bibitem{deRham:2017aoj}
  C.~de Rham and S.~Melville,
  arXiv:1703.00025 [hep-th].


\bibitem{Yoshida:2017swb}
  D.~Yoshida, J.~Quintin, M.~Yamaguchi and R.~H.~Brandenberger,
  arXiv:1704.04184 [hep-th].

%

\bibitem{Misonoh:2016btv}
Y.~Misonoh, M.~Fukushima and S.~Miyashita,
Phys.\ Rev.\ D {\bf 95}, 4, 044044 (2017)
[arXiv:1612.09077 [gr-qc]].

\bibitem{Giovannini:2016jkf}
M.~Giovannini,
Phys.\ Rev.\ D {\bf 95}, 8, 083506 (2017)
[arXiv:1612.00346 [hep-th]].



\bibitem{Horndeski:1974wa}
  G.~W.~Horndeski,
  Int.\ J.\ Theor.\ Phys.\  {\bf 10}, 363 (1974).

\bibitem{Deffayet:2009mn}
  C.~Deffayet, S.~Deser and G.~Esposito-Farese,
  Phys.\ Rev.\ D {\bf 80}, 064015 (2009)
  [arXiv:0906.1967 [gr-qc]].

\bibitem{Kobayashi:2011nu}
  T.~Kobayashi, M.~Yamaguchi and J.~Yokoyama,
  Prog.\ Theor.\ Phys.\  {\bf 126}, 511 (2011)
  [arXiv:1105.5723 [hep-th]].

\bibitem{Gleyzes:2014dya}
  J.~Gleyzes, D.~Langlois, F.~Piazza and F.~Vernizzi,
  Phys.\ Rev.\ Lett.\  {\bf 114}, 21, 211101 (2015)
  [arXiv:1404.6495 [hep-th]].

\bibitem{Langlois:2015cwa}
  D.~Langlois and K.~Noui,
  JCAP {\bf 1602}, 02, 034 (2016)
  [arXiv:1510.06930 [gr-qc]].

\bibitem{Langlois:2015skt}
  D.~Langlois and K.~Noui,
  JCAP {\bf 1607}, 07, 016 (2016)
  [arXiv:1512.06820 [gr-qc]].

\bibitem{Langlois:2017mxy}
  D.~Langlois, M.~Mancarella, K.~Noui and F.~Vernizzi,
  arXiv:1703.03797 [hep-th].



\bibitem{Chamseddine:2014vna}
  A.~H.~Chamseddine, V.~Mukhanov and A.~Vikman,
  JCAP {\bf 1406}, 017 (2014)
  [arXiv:1403.3961 [astro-ph.CO]].



\bibitem{Chamseddine:2016uef}
  A.~H.~Chamseddine and V.~Mukhanov,
  JCAP {\bf 1703}, 03, 009 (2017)
  [arXiv:1612.05860 [gr-qc]].


\bibitem{Khoury:2001wf}
  J.~Khoury, B.~A.~Ovrut, P.~J.~Steinhardt and N.~Turok,
  Phys.\ Rev.\ D {\bf 64}, 123522 (2001)
  [hep-th/0103239].

\bibitem{Lehners:2008vx}
  J.~L.~Lehners,
  Phys.\ Rept.\  {\bf 465}, 223 (2008)
  [arXiv:0806.1245 [astro-ph]].

\bibitem{Kolevatov:2017voe} 
R.~Kolevatov, S.~Mironov, N.~Sukhov and V.~Volkova,
arXiv:1705.06626 [hep-th].



\bibitem{Li:2005fm}
  M.~z.~Li, B.~Feng and X.~m.~Zhang,
  JCAP {\bf 0512}, 002 (2005)
  [hep-ph/0503268].



\bibitem{Koehn:2015vvy}
  M.~Koehn, J.~L.~Lehners and B.~Ovrut,
  Phys.\ Rev.\ D {\bf 93}, 10, 103501 (2016)
  [arXiv:1512.03807 [hep-th]].

\bibitem{Koehn:2013upa}
  M.~Koehn, J.~L.~Lehners and B.~A.~Ovrut,
  Phys.\ Rev.\ D {\bf 90}, 2, 025005 (2014)
  [arXiv:1310.7577 [hep-th]].




\bibitem{Piao:2003zm}
  Y.~S.~Piao, B.~Feng and X.~m.~Zhang,
  Phys.\ Rev.\ D {\bf 69}, 103520 (2004)
  [hep-th/0310206].
   Y.~S.~Piao,
  Phys.\ Rev.\ D {\bf 71}, 087301 (2005)
  [astro-ph/0502343].
 Y.~S.~Piao, S.~Tsujikawa and X.~m.~Zhang,
  Class.\ Quant.\ Grav.\  {\bf 21}, 4455 (2004)
  [hep-th/0312139].



\bibitem{Liu:2013kea}
  Z.~G.~Liu, Z.~K.~Guo and Y.~S.~Piao,
  Phys.\ Rev.\ D {\bf 88}, 063539 (2013)
  [arXiv:1304.6527 [astro-ph.CO]].




\bibitem{Buchbinder:2007ad}
  E.~I.~Buchbinder, J.~Khoury and B.~A.~Ovrut,
  Phys.\ Rev.\ D {\bf 76}, 123503 (2007)
  [hep-th/0702154].

\bibitem{Ijjas:2016tpn}
A.~Ijjas and P.~J.~Steinhardt,
Phys.\ Rev.\ Lett.\  {\bf 117}, 12, 121304 (2016)
[arXiv:1606.08880 [gr-qc]].

\bibitem{Easson:2011zy}
  D.~A.~Easson, I.~Sawicki and A.~Vikman,
  JCAP {\bf 1111}, 021 (2011)
  [arXiv:1109.1047 [hep-th]].

\bibitem{Biswas:2011ar} 
T.~Biswas, E.~Gerwick, T.~Koivisto and A.~Mazumdar,
Phys.\ Rev.\ Lett.\  {\bf 108}, 031101 (2012)
[arXiv:1110.5249 [gr-qc]];
T.~Biswas, A.~S.~Koshelev, A.~Mazumdar and S.~Y.~Vernov,
JCAP {\bf 1208}, 024 (2012)
[arXiv:1206.6374 [astro-ph.CO]].

\bibitem{Odintsov:2014gea}
S.~D.~Odintsov and V.~K.~Oikonomou,
Phys.\ Rev.\ D {\bf 90} (2014) no.12,  124083
[arXiv:1410.8183 [gr-qc]];
S.~D.~Odintsov and V.~K.~Oikonomou,
Phys.\ Rev.\ D {\bf 91} (2015) no.6,  064036
[arXiv:1502.06125 [gr-qc]];
S.~D.~Odintsov and V.~K.~Oikonomou,
Phys.\ Rev.\ D {\bf 92} (2015) no.2,  024016
[arXiv:1504.06866 [gr-qc]];
S.~Nojiri, S.~D.~Odintsov and V.~K.~Oikonomou,
Phys.\ Rev.\ D {\bf 93} (2016) no.8,  084050
[arXiv:1601.04112 [gr-qc]].






\bibitem{Banerjee:2016hom} 
S.~Banerjee and E.~N.~Saridakis,
Phys.\ Rev.\ D {\bf 95}, no. 6, 063523 (2017)
[arXiv:1604.06932 [gr-qc]].


\bibitem{Hendi:2016tiy} 
S.~H.~Hendi, M.~Momennia, B.~Eslam Panah and M.~Faizal,
Astrophys.\ J.\  {\bf 827}, no. 2, 153 (2016)
[arXiv:1703.00480 [gr-qc]];
S.~H.~Hendi, M.~Momennia, B.~Eslam Panah and S.~Panahiyan,
Physics of the Dark Universe 16, 26 (2017)
[arXiv:1705.01099 [gr-qc]].

\bibitem{Sebastiani:2016ras} 
L.~Sebastiani, S.~Vagnozzi and R.~Myrzakulov,
Adv.\ High Energy Phys.\  {\bf 2017}, 3156915 (2017)
[arXiv:1612.08661 [gr-qc]].

\bibitem{Cognola:2016gjy} 
G.~Cognola, R.~Myrzakulov, L.~Sebastiani, S.~Vagnozzi and S.~Zerbini,
Class.\ Quant.\ Grav.\  {\bf 33}, no. 22, 225014 (2016)
[arXiv:1601.00102 [gr-qc]].

\bibitem{Ijjas:2016pad}
A.~Ijjas, J.~Ripley and P.~J.~Steinhardt,
Phys.\ Lett.\ B {\bf 760}, 132 (2016)
[arXiv:1604.08586 [gr-qc]].


\bibitem{Firouzjahi:2017txv}
  H.~Firouzjahi, M.~A.~Gorji and A.~Hosseini Mansoori,
  arXiv:1703.02923 [hep-th].




\bibitem{Hirano:2017zox}
S.~Hirano, S.~Nishi and T.~Kobayashi,
arXiv:1704.06031 [gr-qc].



\bibitem{Zheng:2017qfs}
Y.~Zheng, L.~Shen, Y.~Mou and M.~Li,
arXiv:1704.06834 [gr-qc].





%







\end{thebibliography}
 \end{document}